%% file: main.tex
\documentclass[12pt]{article}
\usepackage{standalone}
\usepackage{import}
\usepackage{amsmath}
\usepackage{graphicx}
\usepackage{enumerate}
\usepackage{natbib}
\usepackage{xurl} % not crucial - just used below for the URL 

\newcommand{\blind}{1}
\newcommand{\fulltitle}{Estimating the duration of RT-PCR positivity for SARS-CoV-2 from doubly interval censored data with undetected infections}

\usepackage[utf8]{inputenc}
\usepackage{amssymb}
\usepackage{floatpag}
\usepackage{bm}
\usepackage[unicode,psdextra]{hyperref}
\usepackage[nameinlink,capitalise,noabbrev]{cleveref}
\usepackage{csquotes}
\usepackage[T1]{fontenc}
\usepackage{textcomp} % provide symbols
\usepackage{xr}
\usepackage{afterpage}
\usepackage{caption}
\usepackage{numprint}
\npfourdigitnosep
\npdecimalsign{.}
\usepackage{sidecap}

\usepackage[font=small,skip=1pt]{caption} 

\def\dist{\sim}

\newcommand{\ind}{\mathrel{\perp\!\!\!\perp}}
\DeclareMathOperator{\prob}{\mathrm{Pr}}

\DeclareMathOperator{\indicator}{\mathbb{I}}

\DeclareMathOperator{\logit}{logit}

\DeclareMathOperator{\Poi}{Poisson}

\DeclareMathOperator{\NBr}{NegBin}
\newcommand{\NBc}{\NBr}

\DeclareMathOperator{\GamDist}{Gamma}
\DeclareMathOperator{\MN}{Multinomial}

\DeclareMathOperator{\MNorm}{N}

\DeclareMathOperator{\expit}{expit}
\newcommand\matr{\bm}
\newcommand\set{\mathcal}
\renewcommand{\vec}[1]{\bm{#1}}
\newcommand{\ssep}{:}

\newcommand{\dmax}{d_\text{max}}
\newcommand{\psens}{p_\text{sens}}
\newcommand{\psenss}{p_\text{sens}^{(s)}}
\newcommand{\psensi}{p_\text{sens}^{(i)}}
\newcommand{\ntot}{n_\text{tot}}
\newcommand{\ndet}{n_\text{d}}
\newcommand{\nnodet}{n_\text{u}}
\newcommand{\pnodet}{p_\text{u}}

\newcommand{\Ncis}{N_\text{CIS}}

\newcommand{\na}{\vec{n}_\text{obs}}

\newcommand{\sched}{\mathcal{T}}

\newcommand{\inform}{{_{\text{inform}}}}
\newcommand{\posResults}{r_{+}}
\newcommand{\negResults}{r_{-}}

\usepackage{enumitem}
\setlist{noitemsep}
\usepackage{xspace}
\makeatletter
\DeclareRobustCommand\onedot{\futurelet\@let@token\@onedot}
\def\@onedot{\ifx\@let@token.\else.\null\fi\xspace}
\def\eg{e.g\onedot} 
\def\ie{i.e\onedot}

\makeatother

\begin{document}

\def\spacingset#1{\renewcommand{\baselinestretch}%
{#1}\small\normalsize} \spacingset{1}

%%%%%%%%%%%%%%%%%%%%%%%%%%%%%%%%%%%%%%%%%%%%%%%%%%%%%%%%%%%%%%%%%%%%%%%%%%%%%%

\if1\blind
{
  \title{\vspace{-1.7cm} \bf \fulltitle}
  \author{%
    Joshua Blake\thanks{%
        MRC Biostatistics Unit, University of Cambridge.
        JB was supported by Bayes4Health (EPSRC  EP/R01856/1).
        JB, PB, and DDA were supported by the UK Medical Research Council (MRC) programme MRC\_MC\_UU\_00002/11. 
        PB, TH, and DDA were supported by the Wellcome Trust (227438/Z/23/Z).
        ASW and KBP were supported by the National Institute for Health Research (NIHR) Health Protection Research Unit in Healthcare Associated Infections and Antimicrobial Resistance at Oxford University in partnership with the UK Health Security Agency (UKHSA) (NIHR200915).
        ASW was supported by the NIHR Biomedical Research Centre, Oxford.
        KBP is supported by the Huo Family Foundation, and the Medical Research Foundation (MRF-160-0017-ELP-POUW-C0909).
        BDMT is supported through the MRC programme grant (MC\_UU\_00002/2) and theme funding (MC\_UU\_0002/20 - Precision Medicine).
        For the purpose of open access, the authors have applied a Creative Commons Attribution (CC BY) license to any Author Accepted Manuscript version arising.
    };
    Paul Birrell\thanks{UK Health Security Agency; MRC Biostatistics Unit, University of Cambridge};
    A.\ Sarah Walker\thanks{Nuffield Department of Medicine, University of Oxford; NIHR Oxford Biomedical Research Centre; NIHR Health Protection Research Unit in Healthcare Associated Infections and Antimicrobial Resistance, University of Oxford.};
     Koen B.\ Pouwels\thanks{Nuffield Department of Primary Care Health Sciences, University of Oxford; National Institute for Health Research Health Protection Research Unit (NIHR HPRU) in Healthcare Associated Infections and Antimicrobial Resistance at the University of Oxford}; \\
     Thomas House\thanks{Department of Mathematics, University of 
Manchester};
     Brian D M Tom\thanks{MRC Biostatistics Unit, University of Cambridge. };
     Theodore Kypraios\thanks{School of Mathematical Sciences, University of Nottingham};
     and Daniela De Angelis\thanks{MRC Biostatistics Unit, University of Cambridge; UK Health Security Agency}
  }
  \maketitle
} \fi

\if0\blind
{
  \bigskip
  \bigskip
  \bigskip
  \begin{center}
    {\LARGE\bf \fulltitle}
  \end{center}
  \medskip
} \fi
% \medskip
\vspace{-1cm}

\begin{abstract}
Monitoring the incidence of new infections during a pandemic is critical for an effective public health response. General population prevalence surveys for SARS-CoV-2 can provide high-quality data to estimate incidence.  However, estimation relies on understanding the distribution of the duration that infections remain detectable. This study addresses this need using data from the Coronavirus Infection Survey (CIS), a long-term, longitudinal, general population survey conducted in the UK. Analyzing these data presents unique challenges, such as doubly interval censoring, undetected infections, and false negatives. We propose a Bayesian nonparametric survival analysis approach, estimating a discrete-time distribution of durations and integrating prior information derived from a complementary study. Our methodology is validated through a simulation study, including its resilience to model misspecification, and then applied to the CIS dataset. This results in the first estimate of the full duration distribution in a general population, as well as methodology that could be transferred to new contexts.
\end{abstract}

\noindent%
\vfill

\newpage
\section{Introduction} \label{sec:intro}

The most acute phase of the COVID-19 pandemic caused by the SARS-CoV-2 virus (2020 and 2021) killed approximately 30 million people globally~\citep{whoCOVIDExcess}, stretched healthcare services to the brink of collapse~\citep{fongNHS}, and disrupted societies worldwide.
This pandemic highlighted the critical need for an effective public health response, informed by key quantities such as the \emph{incidence of infection}, defined as the rate at which new infections occur, which drives important health outcomes including hospital admissions and deaths. 

Incidence of infection can be estimated through convolution approaches~\citep[e.g.][]{brookmeyerBackcalculation},  which relate observations of disease outcomes with incidence through a delay distribution.
In the context of SARS-CoV-2, this approach requires: a time series of the proportion of the population that have a detectable infection and the distribution of the length of time an infection remains detectable using a particular test.
In the UK, estimates of the proportion of the population that have detectable levels of the virus are available from large-scale prevalence surveys~\citep{cisMethodsONS,rileyREACT}.

Here, we estimate the distribution of the duration of infection episodes.
As SARS-CoV-2 is detected by performing RT-PCR testing on an appropriate biological sample (\eg a nasal swab), this duration is the period over which an infected individual would return a positive RT-PCR result.
A previous meta analysis reported a mean duration of detectability of 14.6 days (95\% CI: 9.3--20.0 days)~\citep{cevikShedding}; however, it was based on studies with short follow-up.
Other estimates include only hospitalized patients and/or had unclear inclusion criteria~\citep{ealesCharacterising,hellewellPCRSensitivity}; hence they may not be representative of the general population.

In this paper, we investigate an alternative source of data for the estimation of the duration of an infection episode, the Coronavirus Infection Survey (CIS), run by the Office for National Statistics (ONS)~\citep{CIS}.
The CIS was a unique, large-scale, longitudinal study of RT-PCR positivity in the general population, testing up to 400,000 individuals for nearly three years (see \cref{sec:data}).
As the testing schedule was independent of infection status and the CIS had very long follow-up period, the study provides a unique opportunity to provide an estimate of the distribution of the duration of an infection episode in the general population.
However, the CIS study design, with testing intervals of up to four weeks, and its implementation pose several methodological challenges.

\begin{figure}
  \makebox[\textwidth][c]{\includegraphics{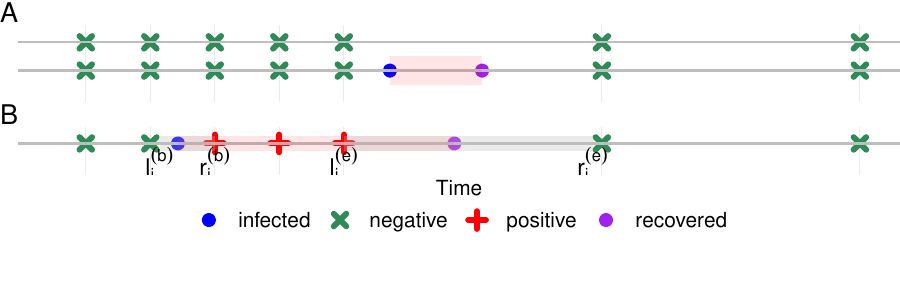}}
  \caption{%
    Challenges posed by the CIS design.
    (A) Undetected episodes.
    (B) Doubly interval censored episodes, shaded regions indicate bounding regions (notation formally defined later).
  }
  \label{fig:challenges}
\end{figure}

Firstly, infections can remain undetected.
Four-weekly testing is longer than the duration of detectability for around two-thirds of infection episodes~\citep{killingleySafety}, and detectability could begin and end between tests.
Specifically, for individuals without a positive test, we do not know if they were infected but undetected, or if they were never infected (compare the two situations visualized in \cref{fig:challenges}(A)).
Also, infection episodes that are detectable for longer are more likely to be detected; hence, detected infection episodes are not a representative sample of all infection episodes.
This selection effect is different from standard left truncation~\cite[e.g.][]{sunEmpirical,bacchettiNonparametric} because here we have information on the individuals in which the short infection episodes occur (\ie their testing times), which constrains undetected infection episode lengths.

Secondly, the duration data are doubly interval censored: we observe the beginning of an episode when a previously negative individual returns a positive test, but the episode could start at any point between those two tests; and the end of the episode is similarly interval censored (see \cref{fig:challenges}(B)).

Thirdly, test results can be misclassified.
The sensitivity, although not the specificity, of the testing procedure is substantially less than 100\%~\citep{cisMethodsONS}.
We refer to clinical sensitivity and specificity throughout this article, incorporating all reasons for misclassified test results, including poor self-swabbing and contamination.
In particular, the negative test providing the upper bound to an episode could be a false negative, incorrectly resulting in a shorter duration.
False negatives may also lead to higher number of undetected episodes.

Fourthly, there is a lack of information on the first 13 days of the duration distribution, which is important to characterize correctly because previous work estimates it contains the distribution's median.
For an individual with perfect adherence to the CIS testing schedule protocol, the most precise information on an infection episode lasting 13 days is provided by a single positive test with negative tests in the preceding and following seven days.
Data from the CIS dataset do not allow us to distinguish whether this infections episode lasted one day, 13 days, or anywhere in-between.

Methods that deal with doubly interval censored data exist~\citep{sunStatistical,bogaertsSurvival}.
However, few consider the additional challenges of undetected episodes;
theoretical frameworks~\citep{turnbullEmpirical,dempsterMaximum} have only applied to the special case where the terminating event is either uncensored or right censored~\citep[e.g.][]{sunEmpirical,bacchettiNonparametric,shenNonparametric}; and
inference in these studies depends on the interval lengths changing negligibly throughout, which does not apply to the CIS where the interval lengths change from one to four weeks. 
\citet{heiseyModelling} generalize the theoretical framework to allow arbitrary patterns of detection times and censoring.
They categorize each combination of possible beginning and end times for an episode into whether the episode would be detected, and whether it is compatible with the pattern of interval censoring observed.
This allows them to build a conditional likelihood, accounting for both these challenges in a context where possible detection times are common to all individuals, leading to a simplification of the likelihood and a common probability of detection across all individuals.
The inclusion of false negatives into survival analysis is an area of much interest, with
\citet{piresIntervalMisclassify} providing a comprehensive review of approaches. However, including false negatives with either doubly interval censored data or missed events has not been addressed. The CIS data require both.

Here, we adopt a Bayesian survival analysis framework to estimate the distribution of infection episode durations from the CIS data.
A Bayesian approach can incorporate prior knowledge to provide information on short episodes~\citep{caoBias}; uncertainty quantification, which is otherwise challenging in this setting~\citep{sunStatistical,dengNonparametric}; and the ability to include the undetected infections through data augmentation.

The paper is structured as follows: we start by describing the design of the CIS and the data it collects in \cref{sec:data}, then develop the survival analysis framework in \cref{sec:modelling}.
In \cref{sec:false-negatives}, we extend the framework to incorporate false negative tests and in \cref{sec:parameters-priors} propose two possible priors for the survival time.
We test the framework in a simulation study in \cref{sec:simulation}, apply it in \cref{sec:CIS}, and conclude with a discussion in \cref{sec:discussion}.

\section{Coronavirus Infection Survey} \label{sec:data}

The CIS~\citep{CIS} was set up in April 2020 and was globally unique in providing a representative, longitudinal, and large-scale study across almost three years of the pandemic.
The study had a household-based design inviting all individuals aged 2 years and over from households randomly selected from previously surveys and address lists.
Once invited and consented, an enrolment swab would be taken at the first visit followed by an optional 4 further weekly visits (giving a total of 5 swabs on days 0, 7, 14, 21, 28 relative to enrolment) after which visits were four-weekly. Furthermore, at each visit, a questionnaire was completed.  In reality, visits were often not on this precise schedule, and occasionally visits were missed.
A full description of the study can be found in the study protocol~\citep{cisProtocol}.
Enrollment was continuously ongoing until 31 January 2022, with data collected until 13 March 2023~\citep{weiRisk}. 

We focus on the period between 10 October 2020 and 6 December 2020 inclusive over which the CIS estimated stable infection prevalence of around 1\%~\citep{onsCISdec2020}.
Stable infection prevalence allows an assumption of constant incidence, simplifying analyses.
Additionally, this period is prior to both vaccination and before the alpha variant was dominant~\citep{lythgoeLineage}, which potentially impacted the duration of infection episodes~\citep{hakkiOnset,russellWithinhost}.
Our analysis includes the cohort of $\Ncis = \numprint{437590}$ CIS participants with at least one test in this period.

We denote by $\sched_i$ the set of times individual labeled $i \in \{ 1, \dots, \Ncis \}$ is tested.
Time is defined such that the first day of the period considered, 10 October 2020, is day 1 and the final day of the period considered, 6 December 2020, is day $T = 58$.
The smallest element of $\sched_i$ is individual $i$'s last test prior to time 1, if it exists, or their first test following enrollment in the study otherwise, and including all subsequent tests even those that occur after day $T$.
Each individual has exactly one test schedule, but it is possible that $\sched_i = \sched_{i'}$ for $i \neq i'$; this occurs commonly for individuals in the same household.
We assume that the test schedules are uninformative on all quantities of interest, in particular the presence or absence of infection, because their timings were prespecified in the study design.
Therefore, we condition on them implicitly in all the calculations that follow.

Positive tests with negatives in-between in the same individual may or may not be classed as the same infection episode using a pre-existing heuristic based on the time between the tests, the number of negative tests between the positives, and the variant of the infection; see \citet{weiRisk} for details.
For the $j$th infection episode, there are up to four important times: the day after the negative test in that individual prior to the first positive test, $l_j^{(b)}$; the day of the first positive test, $r_j^{(b)}$; the day of the last positive test, $l_j^{(e)}$; and the day before the subsequent negative test, $r_j^{(e)}$ (see \cref{fig:challenges}(B)).

We include only episodes for which: (i) the episode's first positive test occurred between 10 October 2020 and 6 December 2020 inclusive, \ie $r_j^{(b)} \in [1, T]$; and (ii) a negative test bounds both the beginning and end of the episode,
therefore, both $l_j^{(b)}$ and $r_j^{(e)}$ exist.
In total, there were $\ndet = \numprint{4800}$ such detected episodes.

\section{A Bayesian estimation approach} \label{sec:modelling}

The target of inference is $\vec{\theta}$, the parameters of the survival function $S_{\vec{\theta}}(t) = \prob(D_j \geq t \mid \vec\theta)$ for the random variable $D_j$, representing the number of days on which a positive result would be returned by a RT-PCR testing procedure that has 100\% specificity and sensitivity due to infection episode $j$.
Here, 100\% an individual's true infection status refers to whether their viral load is above the test's limit of detection.
We adopt a Bayesian framework to derive a posterior distribution for $\vec{\theta}$ given appropriate prior knowledge (\cref{sec:parameters-priors}) and partial information provided by the set of observed vectors $\{ o_j \}$ (defined later), accounting for the doubly interval censoring and undetected episodes.
In what follows, we explain the data generating process and derive a statistical model for the case where that there are no misclassified test results; in \cref{sec:false-negatives} we generalize the framework to include false negatives.

An important aspect of our approach is that the dimension of the posterior distribution does not increase with either the cohort size of number of detected infections.
Previous approaches to related problems augmented the data with a parameter per detected infection~\citep{heBayesiana,heBayesian,caoModeling}.
Such an approach would be computationally prohibitive here because of the large number of detected infections and the regulatory requirement for the data to be stored on a Trusted Research Environment, the ONS Secure Research Service (SRS)~\citep{onsSRS}, which means that methods requiring high-performance computing cannot be used.

\subsection{Data generating process}

For our purposes, the $j$th ($j = 1, \dots, \ntot$, where $\ntot$ is the unobserved total number of infection episodes) infection episode is the triplet $W_j = (B_j, E_j, i_j)$ where $B_j$ is the beginning of the episode, the first day the individual is detectable; $E_j$ is the end of the episode, the last day the individual is detectable; and $i_j$ is the individual in which the $j$th infection occurs.
The episode's duration is a deterministic transformation of these variables, $D_j = E_j - B_j + 1$; and we assume that $B_j$, $D_j$, and $i_j$ are independent.
The independence of infection times is a simplifying assumption that does not hold in a household-based survey such as CIS but is required for the problem to remain analytically tractable; independence is discussed further in \cref{sec:discussion}.
$D_j$'s distribution is determined by $\vec{\theta}$; both $B_j$ and $i_j$ are modelled as draws from uniform distributions over their respective discrete state spaces.
Additionally, we assume that the $D_j$s are independent of each other and $B_j$ (\ie the length of the episode is independent of when it occurs), and identically distributed.

To make this problem tractable, we model infection episodes as independent conditional on $\ntot$ and $\vec{\theta}$; \ie $W_j \ind W_{j'} \mid \ntot, \vec{\theta}$ for $j \neq j'$.
This is true both for infection episodes within the same individual and across individuals; this assumption is discussed in \cref{sec:discussion}.

The test results for any individual is fully determined by any infection episodes that occur in that individual and their test schedule.
The set of times at which individual $i_j$ tests positive at due to $W_j$ is $\posResults(W_j) = \{ t \in \sched_{i_j} \ssep B_j \leq t \leq E_j \}$.
The negative test immediately before the start of the episode (if there is one) is $\negResults^{(b)}(W_j) = \max \{ t \in \sched_{i_j} \ssep t < B_j \}$.
The negative test immediately after the end of the episode (if there is one) is $\negResults^{(e)}(W_j) = \min \{ t \in \sched_{i_j} \ssep t > E_j \}$.

Next, we define $O_j$ to contain the test results due to $W_j$ that give information on $W_j$'s length, and hence are relevant to inference about $\vec{\theta}$.
If $\posResults(W_j) = \varnothing$, $\min \posResults(W_j) \notin [1, T]$, $\negResults^{(b)}$ is undefined, or $\negResults^{(e)}$ is undefined then the episode is undetected, in which case $O_j = \varnothing$.
Otherwise, $O_j = [l_j^{(b)}, r_j^{(b)}, l_j^{(e)}, r_j^{(e)}, i_j]^T$, where $i_j$ is the individual in which the episode occurs; $l_j^{(b)} = \negResults^{(b)}(W_j) + 1$ and $r_j^{(b)} = \min \posResults(W_j)$ are the earliest and latest time the episode could have begun and produced the observed series of test results respectively; and $l_j^{(e)} = \max \posResults(W_j)$ and $r_j^{(e)} = \negResults^{(e)}(W_j) - 1$ are the earliest and latest time the episode could have ended respectively.
These definitions coincide with the way the dataset is constructed (see \cref{sec:data}).

Ignoring misclassified test results, the latent time $B_j$ for a detected infection episode $j$ is bounded between, $l_j^{(b)}$, the day after the last negative test; and $r_j^{(b)}$, the day of the first positive test associated with episode $j$ (see \cref{fig:challenges}(B)).
Similarly, $E_j$ is bounded by $l_j^{(e)}$, the day of the last positive test; and $r_j^{(e)}$, the day before the following negative test.
That is $B_j \in [l_j^{(b)}, r_j^{(b)}]$ and $E_j \in [l_j^{(e)}, r_j^{(e)}]$.
Therefore, all information from the observations of a detected infection episode $j$ is fully contained in the vector $O_j$.

\subsection{Set up} \label{sec:inference}

\begin{figure}
    \makebox[\textwidth][c]{\includegraphics[width=0.9\paperwidth]{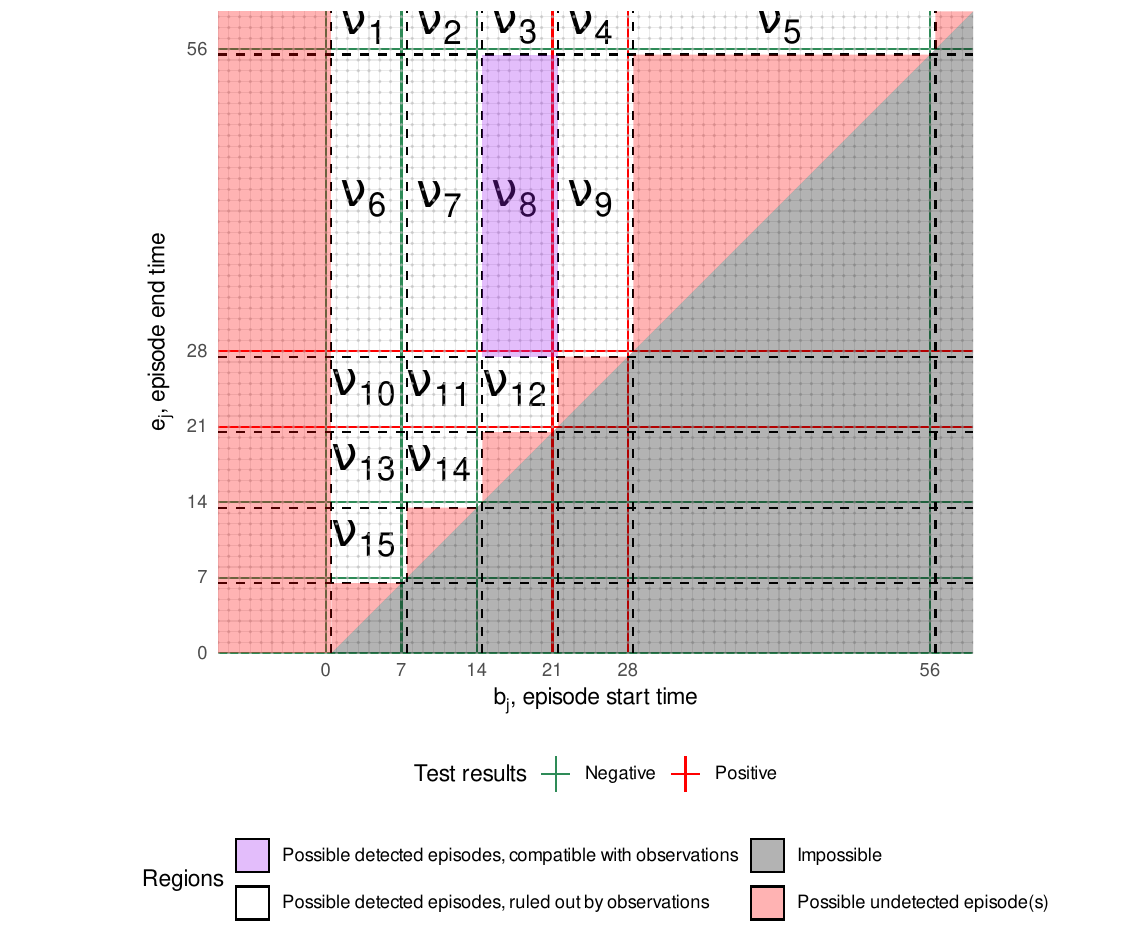}}
    \thisfloatpagestyle{empty}
    \caption[Episode regions]{%
      Each dot is a combination of $b_j$ and $e_j$ for an arbitrary individual $i$.
      The combinations giving rise to the same value of $\nu_k$ are in the same box, bounded by dashed lines.
      $i$ had negative tests at times 0, 7, 14, 56, and 84 (not shown) and positive tests at times 21 and 28.
      The purple region corresponds to a doubly interval censored episode observed in this individual.
      That is, $n_8 = 1$ and $n_k = 0$ for $k = 1, \dots, 7, 9, \dots, 15$.
      The red region corresponds to combinations giving $O_j = \varnothing$.
      The grey impossible region violates $b_j \leq e_j$.
    }
    \label{perf-test:fig:partitionSpace}
\end{figure}

Define an integer $N_E = |\set{E}|$ (the cardinality of $\set{E}$) and $\set{E} = \{ \vec{\nu}_1, \dots, \vec{\nu}_{N_E} \}$ as the set of all possible values $O_j$ can take, excluding $\varnothing$, conditional on the test schedules but no other data; that is, $O_j \in \set{E}$ if and only if $j$ is a detected infection.
Let $\vec{\nu}_k = [l^{(b)}_k, r^{(b)}_k, l^{(e)}_k, r^{(e)}_k, i_k]^T$ be an arbitrary member of $\set{E}$, visualized in \cref{perf-test:fig:partitionSpace}.

Let $n_k$ denote the number of times that $\vec{\nu}_k$ appears in the episodes dataset (\ie the observed data); $\nnodet$ denote the latent number of undetected episodes; and $\vec{n} = [n_1, \dots, n_{N_E}, \nnodet]^T$.
Hence, $\ntot = \ndet + \nnodet = \sum_{i=1}^{N_E} n_i + \nnodet$.
If $\vec{n}$ was known, then the problem reduces to the well-studied problem of inferring a distribution from doubly interval censored data; however, only $\na = [n_1, \dots, n_{N_E}]^T$ is observed.
Therefore, to infer $\vec\theta$, we augment $\na$ with the latent quantity $\nnodet$.

Let $p_k = \prob(O_j = \vec{\nu}_k \mid \vec{\theta})$, the probability that $O_j$ takes the value $\vec{\nu}_k$ for $k = 1, \dots, N_E$.
Similarly, let $p_u = \prob(O_j = \varnothing \mid \vec{\theta})$, the probability that $j$ is undetected.
Then, the probability distribution for $O_j$ is specified by $\vec{p} = [p_1, \dots, p_{N_E}, p_u]^{\mathsf{T}}$.

As we assume independence:
\begin{align}
  \vec{n} \mid \ntot, \vec{\theta} &\dist \MN(\ntot, \vec{p})
\intertext{that is:}
  p(\vec{n} \mid \ntot, \vec{\theta}) &= \frac{\ntot!}{\nnodet!\prod_{k=1}^{N_E} n_k!} p_u^{\nnodet} \prod_{k=1}^{N_E} p_k^{n_k}.
  \label{perf-test:eq:multinomial-ll}
\end{align}

In the CIS data, each $n_k$ ($k \neq u$) is observed as either 0 or 1.
Define $\set{D} = \{ k \ssep n_k = 1 \}$, the set of detected episodes.
Furthermore, note that the support of the multinomial distribution requires that $\nnodet = \ntot - \ndet$.
Then \cref{perf-test:eq:multinomial-ll} simplifies to:
\begin{align}
  p(\vec{n} \mid \ntot, \vec{\theta})
  &= p(\na \mid \ntot, \vec{\theta}) \\
  &= \frac{\ntot!}{(\ntot - \ndet)!} p_u^{\ntot-\ndet} \prod_{k \in \set{D}} p_k.
  \label{perf-test:eq:multinomial}
\end{align}

The relevant information from the CIS data is fully contained in the vector $\na$.
Therefore, the posterior of interest is (full derivations of this and subsequent quantities are in \cref{sec:derivations}):
\begin{align}
p(\vec{\theta} \mid \na)
&\propto  p(\vec{\theta}) \left( \prod_{k \in \set{D}} p_k \right) \left( \sum_{\ntot=\ndet}^\infty p(\ntot \mid \vec{\theta}) \frac{\ntot!}{(\ntot - \ndet)!} \pnodet^{\ntot - \ndet} \right).
\label{perf-test:eq:posterior1}
\end{align}

For mathematical convenience, we assume the prior $\ntot \dist \NBc(\mu, r)$ (where $\mu$ is the prior mean and $r$ its overdispersion), and that it is independent of $\vec{\theta}$, the parameters of the survival distribution.
In this case \cref{perf-test:eq:posterior1} simplifies to:
\begin{align}
p(\vec{\theta} \mid \na)
&\propto p(\vec{\theta}) \left( \prod_{i \in \set{D}} p_k \right) (r + \mu (1- \pnodet))^{-(r+\ndet)} \label{perf-test:eq:full-posterior}.
\end{align}
The rest of this section derives expressions for $p_{k}$ and $p_{u}$.

Decompose $p_k$ as $p_k = p_{ik} \prob(i_j = i_k \mid \vec{\theta})$
where $p_{ik} = \prob(O_j = \vec{\nu}_k \mid i_j = i_k, \vec{\theta})$.
Assuming all individuals are equally likely to be infected, $\prob(i_j = i_k \mid \vec{\theta}) = 1/\Ncis$ for all $j$ and $k$.

Therefore, $p_{ik}$ takes the standard form of the likelihood for doubly interval censored data without truncation~\citep[e.g.][]{sunEmpirical}:
\begin{align}
p_{ik}
\propto& \sum_{b = l_k^{(b)}}^{r_k^{(b)}} \left( S_{\vec{\theta}}(l_k^{(e)} - b + 1) - S_{\vec{\theta}}(r_k^{(e)} - b + 2) \right).
\label{perf-test:eq:pia}
\end{align}

The remaining component of \cref{perf-test:eq:full-posterior} required is $1- p_u$, one minus the probability of missing an infection, \ie the probability of detecting an infection.

\subsection{Deriving $1 - p_u$} \label{sec:prob-undetected}

The final component of \cref{perf-test:eq:full-posterior} required is $1 - p_u$.
\Cref{sec:derivations} shows that
\begin{align}
  1 - p_u
  &= \frac{1}{\Ncis} \sum_{i=1}^{\Ncis} (1 - \prob(O_j = \varnothing \mid i_j = i, \vec{\theta})).
  \label{perf-test:eq:pu}
\end{align}
Let $p_{iu} = \prob(O_j = \varnothing \mid i_j = i, \vec{\theta})$.
An episode $j$ in individual $i_j$ is detected if and only if all the following conditions are met.
\begin{enumerate}
    \item 
    $t\in[B_j, E_j]$ for some $t \in \sched_{i_j}$;
    \ie there is at least one positive test for the episode.
    \item $B_j > \min ( \sched_{i_j} )$.
      For individuals enrolled during the period considered ($\min \sched_{i_j} > 0$), this ensures that the beginning of the episode is lower bounded.
      For individuals enrolled prior to the period considered ($\min \sched_{i_j} \leq 0$), this means that the episode was not detected prior to time 1.
    \item $B_j \leq T_{i_j}$ where $T_{i_j} = \max \{ t \in \sched_{i_j} \ssep t \leq T \}$ is the last time that $i_j$ is tested in the period, meaning that the test is detected within the period.
    \item $\exists t \in \sched_{i_j}$ such that $t > E_j$, upper bounding the end of the episode.
      For episodes detected in the period we consider, a negligible number of episodes are excluded due to this criteria (<5\% of detected episodes, themselves likely around 15\% of all episodes).
      Therefore, we assume this occurs with probability 0.
\end{enumerate}

First define $\tau_{\sched_i}(t)$ as the time until the next test at or after time $t$ in the schedule $\sched_i$:
\begin{align}
\tau_{\sched_i}(t) &= \min \{ t' \in \sched_i : t' \geq t \} - t
\label{perf-test:eq:tau-def}
\end{align}
The first condition can now be expressed as $e_j \geq b_j + \tau_{\sched_{i_j}}(b_j)$.
Equivalently, $d_j \geq \tau_{\sched_{i_j}}(b_j) + 1$.
Therefore, omitting the conditioning on $\vec{\theta}$ and $i_j = i$:
\begin{align}
1 - p_{iu}
&= \prob(D_j \geq \tau_{\sched_{i}}(B_j)+ 1, \min \sched_{i} < B_j \leq T_{i}) \\
&= \sum_{b = \min \sched_{i} + 1}^{T_{i}} \prob(D_j \geq \tau_{\sched_{i}}(b) + 1 \mid B_j = b) \prob(B_j = b)\\
&\propto \sum_{b = \min \sched_{i} + 1}^{T_{i}} S_{\vec{\theta}}(\tau_{\sched_{i}}(b) + 1).
\label{perf-test:eq:piu}
\end{align}
Substituting into \cref{perf-test:eq:pu}:
\begin{align}
1 - p_u
& \propto \sum_{i=1}^{\Ncis} \sum_{b = \min \sched_{i} + 1}^{T_{i}} S_{\vec{\theta}}(\tau_{\sched_{i}}(b) + 1).
\end{align}
For computational efficiency, note that this can be rewritten as:
\begin{align}
  1- p_u
  &\propto \sum_{t=1}^{\dmax} S_{\vec\theta}(t) m_t
  \label{eq:Sm-sum} \\
  m_t &= \sum_{i=1}^{\Ncis} \sum_{b = \min \sched_{i} + 1}^{T_{i}} \indicator(\tau_{\sched_{i}}(b) + 1 = t)
\end{align}
where $\indicator$ is the indicator function taking the value $1$ when the statement in its argument is true and $0$ otherwise.
The $m_t$s rely only on the test schedules, which are fixed, and can be computed once and stored.
Furthermore, the sum in \cref{eq:Sm-sum} can be efficiently implemented as a dot product.

\section{Handling false negatives} \label{sec:false-negatives}

Now we modify the survival framework to incorporate false negatives by assuming a test sensitivity, $\psens < 1$, but continuing to assume a negligible probability of false positives.
Using a simple model, in particular a constant test sensitivity, means that calculating the likelihood remains tractable.
Additionally, we limit the set of permutations of false negative and true positive tests that we consider to have positive probability.
Specifically, we assume that, if $i_j$ is tested during infection episode $j$, either: the individual has a single false negative test and the episode ends before their next test, or the first test is a true positive test.
This assumption is reasonable because false negatives normally occur when the viral load is low, which is normally late in the infection episode.
For the same reason, we assume there is at most one false negative test following the final true positive test in the episode.
In \cref{sec:simulation}, we show that these assumptions still give acceptable performance as long as $\psens$ is not too low.

The test results which bound $B_j$ and $E_j$, contained in $O_j$, are now random because there could be additional false negative results.
Additionally, all results between these times are random and hence should enter into the likelihood.
For tractability, we consider only tests between the negative tests providing an upper bound on the length of episode $j$.

To do so, define a random vector $O'_j$ with state space $\{\varnothing\} \cup \set{E}'$ which will replace $O_j$.
As before, $O'_j = \varnothing$ if episode $j$ is undetected.
Otherwise, $O'_j \in \set{E}'$ where $\set{E}' = \{ \vec{\nu}_1', \dots, \vec{\nu}_{N_E'}' \}$ augments the space $\set{E}$ with test results during the episode.
The elements of $\set{E}'$ are all $\vec{\nu}_k' = [\vec{\nu}_k, \vec{y}_k]^T$ where $\vec{\nu}_k \in \set{E}$ and $\vec{y}_k$ is a vector of test results for the relevant testing times excluding the negative at $l^{(b)}_k$, $\sched'_k = \{ t \in \sched_{i_k} \ssep r_k^{(b)} \leq t \leq r_k^{(e)} + 1 \}$.
To formalize the definition of $\vec{y}_k$, let $m_k$ denote the size of $\sched'_k$ and denote the elements of $\sched'_k$ by $t_{k,1} < \dots < t_{k,m_k}$.
Then, $\vec{y}_k \in \{0, 1\}^{m_k}$ and satisfies the following two conditions.
\begin{enumerate}
  \item The elements of $\vec{y}_k$ corresponding to the tests at times $r_k^{(b)}$ and $l_k^{(e)}$ are positive, \ie $y_{k,1} = y_{k,m_k-1} = 1$.
  \item The element of $\vec{y}_k$ corresponding to the test at time $r_k^{(e)} + 1$ is negative, \ie $y_{k,m_k} = 0$.
\end{enumerate}
These conditions are due to the construction of the intervals as positive and negative tests bounding the beginning and end times of the episode.

Similarly, we define $p_k'$, $p_{ik}'$, $p_u'$, and $p_{iu}'$ to replace $p_k$, $p_{ik}$, $p_u$, and $p_{iu}$ respectively.
\begin{align}
    p_{ik}' &= \prob(O'_j = \vec{\nu}'_k \mid i_j = i_k, \vec{\theta}) \\
    p_k' &= \frac{1}{\Ncis} p_{ik}' \\
    p_{iu}' &= \prob(O'_j = \varnothing \mid i_j = i, \vec{\theta}) \\
    p_u' &= \frac{1}{\Ncis} \sum_{i=1}^{\Ncis} p_{iu}'.
\end{align}

\subsection{Deriving $p'_{ik}$} \label{imperf-test:sec:modifying-p_ia}

We will modify $p_{ik}$ to form $p_{ik}' = \prob(O'_j = \vec{\nu}'_k \mid i_j = i_k, \vec{\theta})$, taking into account false negatives.
We consider a mixture of two scenarios, defined by whether the final test in $\sched'_{k}$ is a false negative, with the mixture probability determined by the test sensitivity.

Similar likelihoods have previous appeared in the literature~\citep[e.g.][eq.\ (2)]{piresIntervalMisclassify}, but for singly interval censored data.
Incorporating the doubly interval censored nature of the CIS data involves summing over the possible episode start times.

By assumption, the negative test bounding the start of the episode, on day $l_k^{(b)}-1$, is a true negative.
True negatives occur with probability 1, and hence this test does not contribute to the likelihood.

As we assume that there are no false positives, the infection episode must span at least the period $[r^{(b)}_k, l^{(e)}_k]$, a period starting and ending with a positive test.
This includes all $t \in \sched'_k$ except $t = r_k^{(e)}+1$.
Therefore, the test results $\vec{y}_k$, except the test at $r_k^{(e)}+1$, are either true positives or false negatives.
This gives $t_+ = \sum_{l=1}^{m_k-1} y_{k,l}(t)$ true positives and $f_- = \sum_{l=1}^{m_k-1} (1 - y_{k,l}(t))$ false negatives.

Consider the negative test at $r_k^{(e)}+1$, the first negative after the start of the episode which may be either a true or false negative.
It is a false negative if and only if the episode ends at or after the test, \ie $E_j > r_k^{(e)}$.
By considering the case of whether this occurred or not and assuming $\psens$ is known and fixed, in \cref{sec:p-ia-dash} we show
\begin{align}
p_{ik}'
&\propto \sum_{b = l_k^{(b)}}^{r_k^{(b)}} S_{\vec{\theta}}(l_k^{(e)} - b + 1) - p_\text{sens} S_{\vec{\theta}}(r_k^{(e)} - b + 2).
\label{imperf-test:eq:pia-prime-constant}
\end{align}
Note that if $p_\text{sens} = 1$ then $p_{ik}' = p_{ik}$ (see \cref{perf-test:eq:pia}).

\subsection{Deriving $p'_{iu}$} \label{imperf-test:sec:modifying-p_iu}

We now modify $p_{iu}$ to form $p_{iu}' = \prob(O'_j = \varnothing \mid i_j = i, \vec{\theta})$ to take into account false negatives.
Several mechanisms for episodes being undetected were previously considered when deriving $p_{iu}$, we now consider the additional mechanisms arising due to false negatives.
Specifically, episode $j$ could be undetected if the first test after $b_j$ is a false negative and then there are no subsequent positive tests.

This false negative would occur at the first test after the infection episode begins, on day $b_j + \tau_{\sched_{i_j}}(b_j)$.
The episode has not yet ended at the time of the test if $e_j = b_j + d_j - 1 \geq b_j + \tau_{\sched_{i_j}}(b_j)$, that is the duration of the infection $d_j \geq \tau_{\sched_{i_j}}(b_j) + 1$.
Conditional on the episode having not yet ended, the test result is negative with probability $1 - \psens$.

For there to be no subsequent positive tests, all tests up until day $e_j$ are false negatives.
By assumption, there is a negligible probability of missing an episode due to two false negative tests.
This is because that would require both a long episode, encompassing two test times, and for both these tests to be false negatives.
Therefore, an episode is undetected only if the episode ends before a second test.
Denote the number of days between $b_j$ and the test following the false negative as $\tau^2_{\sched_{i_j}}(b_j) \stackrel{\text{def}}{=} \tau_{\sched_{i_j}}(\tau_{\sched_{i_j}}(b_j) + 1)$.
The episode ends before this test if $d_j \leq \tau^2_{\sched_{i_j}}(b_j)$.

Therefore, this mechanism causes episode $j$ to be undetected if all the following conditions hold.
\begin{enumerate}
    \item The episode would have been detected considering only the mechanisms in \cref{sec:prob-undetected}. That is $\min(\sched_{i_j}) < b_j \leq T_{i_j}$ and $e_j \geq \tau_{\sched_{i_j}}(b_j) + b_j$.
    \item The episode ends in the interval $[\tau_{\sched_{i_j}}(b_j) + b_j, \tau^2_{\sched_{i_j}}(b_j) + b_j - 1]$.
      Note that the lower bound here is exactly the bound on $e_j$ in the previous condition.
      Equivalently, $\tau_{\sched_{i_j}}(b_j) + 1 \leq d_j \leq \tau^2_{\sched_{i_j}}(b_j)$.
    \item A false negative occurs on day $\tau_{\sched_{i_j}}(b_j) + b_j$. Conditional on the previous condition, this occurs with probability $1 - \psens$.
\end{enumerate}

In \cref{sec:p-iu-dash} we show that this gives:
\begin{align}
1 - p_{iu}'
=& \frac{1}{T} \sum_{b=\min(\sched_{i}) + 1}^{T_{i}} \left( p_\text{sens} S_{\vec{\theta}}(\tau_{\sched_{i}}(b) + 1) + (1 - p_\text{sens}) S_{\vec{\theta}}(\tau^2_{\sched_{i}}(b) + 1)\right).
\label{imperf-test:eq:pit-prime}
\end{align}

\section{The survival function} \label{sec:parameters-priors}

\begin{figure}
  \includegraphics{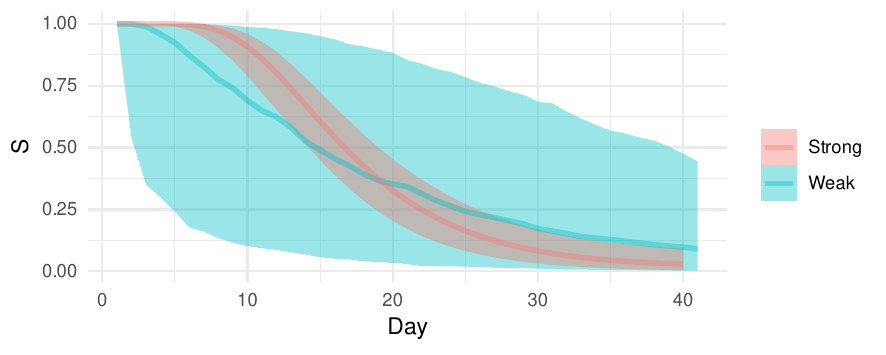}
  \caption{%
    Prior predictive values of $S_{\vec{\theta}}$ for the two priors.
  }
  \label{fig:priors}
\end{figure}

Next we specify the form and priors for $S_{\vec{\theta}}(t)$.
We parameterize $S$ in terms of the discrete-time hazard on day $i$, $\lambda_i = \prob(B = i \mid B \geq i)$, giving $S_{\vec{\theta}}(t) = \prod_{i=1}^{t-1} (1 - \lambda_{i})$; the lack of monotonicity or sum constraint on the hazard makes it an attractive parameterization for inference~\citep{heBayesian}.
Therefore, $\vec{\theta} = [\lambda_1, \dots, \lambda_{\dmax-1}]^{\mathsf{T}}$, where $\dmax$ is the longest possible duration assumed as the maximum possible duration of the observed episodes, \ie $\dmax = \max_{k \in \set{D}} r^{(e)}_k - l^{(b)}_k + 1$.
We consider two priors (depicted in \cref{fig:priors}), with varying degrees of informativeness regarding $S_{\vec{\theta}(t)}$.

The first prior for $S_{\vec{\theta}}$ is weakly informative, centered on prior estimates.
Specifically, we assume an independent prior distribution for each $\lambda_t$ of Beta(0.1, 1.9).
This distribution has mean 0.05, little information (standard deviation of 0.13), and a central 95\% probability mass of 0.00--0.47.
The mean is in line with previous estimates of the median duration~\citep{cevikShedding}.

The second is a strongly informative prior; it incorporates prior information from reliable estimates of $\lambda_t$ for $t < 20$.
We take a previous Bayesian analysis~\citep{blakeThesis} of data from The Assessment of Transmission and Contagiousness of COVID-19 in Contacts (ATACCC) study~\citep{hakkiOnset}, which tested individuals who had been exposed to infection daily up to a maximum of 20 days.
This ATACCC-based analysis produces posterior estimates of $\lambda_t$ with a posterior distribution with positive correlation between $\lambda_t$ and $\lambda_{t'}$, especially for small $|t-t'|$.
Furthermore, the uncertainty in the prior estimates for $\lambda_t$ for $t\geq20$ are underestimated because they are based on extrapolation of the ATACCC data under strong model assumptions.
The following prior, based on the discrete Beta process prior~\citep{ibrahimBayesian,sunStatisticala}, incorporates both these aspects:
\begin{align}
  \logit \vec{h} &\dist \MNorm(\vec{\mu}_A, \matr{\Sigma}_A) \\
  \lambda_t &\dist \text{Beta}(\alpha_t, \beta_t) &t = 1, 2, \dots \\
  \alpha_t &= k_t h_t + \alpha_0 \\
  \beta_t &= k_t (1 - h_t) + \beta_0
\end{align}
where $k_t$, $\alpha_0$, and $\beta_0$ are hyperparameters; and $\vec{\mu}_A$ and $\matr{\Sigma}_A$ are posterior approximations of the ATACCC-based posterior (see \cref{sec:ataccc-prior} for details).
We use $\alpha_0 = 0.1$ and $\beta_0 = 1.9$ to match the weakly informative prior and the following form for $k_t$:
\begin{align}
k_t = \begin{cases}
  \expit(-0.4 \times (t - 20)) &\text{for $t \leq 39$} \\
  0 &t > 39.
\end{cases}
\end{align}
The form and choice of constants reflects the subjective belief that $h_t$ is a good estimate of $\lambda_t$ for small $t$ but increasingly unreliable; specifically, it is large at 0, when ATACCC is reliable, but becomes small for $t \geq 20$.

\section{Simulations} \label{sec:simulation}

We use a simulation study to evaluate the performance of the method and the impact of the simplifying assumptions made.
We simulate avoiding the independence assumption made in \cref{sec:modelling} and without assuming that some patterns of test results have negligible probability, as in \cref{sec:false-negatives}.
Therefore, the model used to generate the data is more realistic than what we will use for inference, testing the impact of these simplifying assumptions.
Further, we consider the impact of misspecifying $\psens$ as this parameter cannot be inferred within our set-up.
We show that, as long as $\psens$ is not too small, inference performance remains acceptable.

\subsection{Setup}

We simulate a dataset of detected episodes that has the same characteristics as that in the CIS by the following procedure.
\begin{enumerate}
    \item Extract the test schedules for each individual who had at least one test during the period of interest.
    \item Draw an episode start time, $b_{j}$ for each individual uniformly at random between 2 July 2020 (100 days before the period where a detected episode would be included) and 6 December 2020 (the end of this period).
    \item Draw a duration of episode for their episode, $d_j$, based on a combination of previous estimates (described in \cref{sec:simulation-truth}). Then calculate the end of their infection episode, $e_{j} = b_{j} + d_i - 1$.
    \item Simulate the test results based on the test schedule, $b_{j}$, and $e_{j}$. A test on day $t$ between $b_{j}$ and $e_{j}$ (inclusive) is positive with probability $\psens$, where $\psens$ can vary with the time since infection, as defined below. All tests outside this interval are negative.
    \item Discard episodes where there are no positive tests (\ie undetected episodes) and then apply the inclusion criteria from \cref{sec:data}. Denote by $p$ the proportion of episodes that are retained.
    \item Of these remaining episodes, sample $\ndet = 4800$ to match the sample size of the true dataset. This is needed because in step 2 the entire cohort was infected, while in the real study only a (unknown) portion is infected.
    \item For this final set of episodes, calculate $(l_j^{(b)}, r_j^{(b)}, l_j^{(e)}, r_j^{(e)})$ by taking the day after the last negative prior to any positives, the first positive, the last positive, and the day before the negative following the last positive respectively.
\end{enumerate}

We simulate four scenarios for the test sensitivity.
The first three are constant, $\psens \in \{ 0.6, 0.8, 1.0 \}$.
The final scenario is a varying test sensitivity, which is more realistic~\citep{blakeThesis}.
Specifically we use the following form:
\begin{equation}
  v(t) = \begin{cases}
    0.9 - \frac{0.9-0.5}{50}t &t \leq 50 \\
    0.5 &t > 50
  \end{cases}
  \label{imperf-test:eq:variable-test-sensitivity}
\end{equation}
where $t$ is the number of days since the infection occurred.
We denote by $\psenss$ the true test sensitivity used in the simulation, with $\psenss = v$ indicating the varying test sensitivity.

For each scenario, we infer the survival function using the procedure proposed in this paper, with a point prior of $\psens$ of 0.6, 0.8, or 1.0.
That is, the value of $\psens$ used in inference is not necessarily $\psenss$, and we consider the impact of this misspecification.
We denote by $\psensi$ the assumed test sensitivity in inference.

If $\psenss = \psensi$, we refer to $\psens$ as being correctly specified; otherwise we refer to it as misspecified.
Note that if $\psenss = v$, then $\psens$ is always misspecified.
Even in the correctly specified case, there is still some misspecification of the model to false negatives owing to the simplifying assumptions made.
The amount that the simplifying assumptions are violated increases as $\psens$ decreases.

Simulation was performed in R~4.2.0~\citep{R-4-2-0} using tidyverse~2.0.0~\citep{tidyverse}.
Inference was implemented in Stan via RStan~2.21.8~\citep{rstan2-21-8} using default settings.
Convergence was assessed using Rhat and ESS~\cite{vehtariRhat}, and all runs checked for divergent transitions.

We used a vague prior for $\ntot$, with $\mu = n_d / p$ and $r = 1$.

\subsection{Results}

\begin{figure}
  \includegraphics[width=\textwidth]{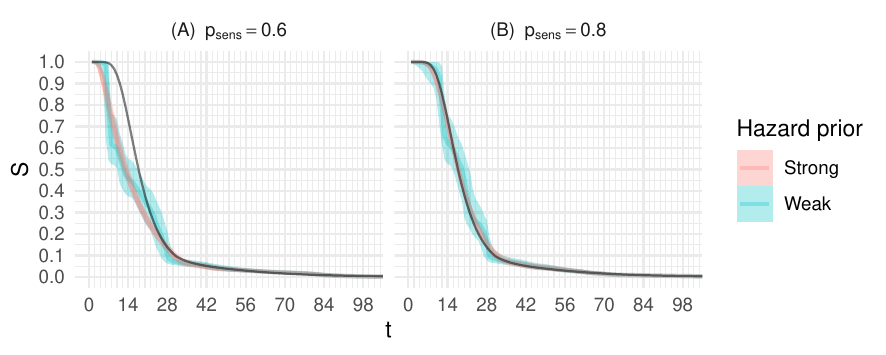}
  \caption[Simulation study results with constant test sensitivity]{%
    Posterior (median and 95\% credible interval (CrI)) survival time for the simulation study with a correctly specified test sensitivity.
    True survival time shown in black.
  }
  \label{imperf-test:fig:constant-test-sensitivity}
\end{figure}

When $\psens = 0.8$ and is correctly specified, the model recovers the true survival time well (see \cref{imperf-test:fig:constant-test-sensitivity}(B)).
The strongly informative prior for $\vec\theta$, in comparison to the weakly informative prior, helps to overcome the misspecification due to the simplifying assumptions, moving the estimated survival function closer to its true survival time; in particular, the central estimate is smoother and has less uncertainty.
However, when $\psens = 0.6$, both priors lead to underestimation (see \cref{imperf-test:fig:constant-test-sensitivity}(A)).
This is likely caused by too large a violation of the simplifying assumptions made in \cref{sec:false-negatives}.

\begin{figure}[ht!]
  \includegraphics[width=\textwidth]{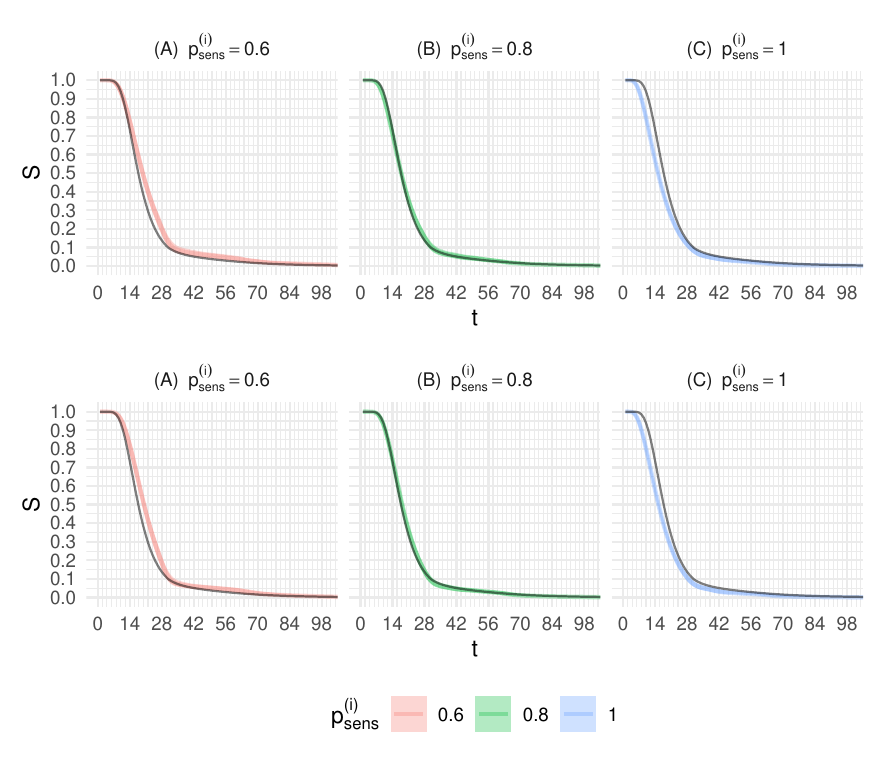}
  \caption[Simulation study results with varying test sensitivity]{%
    Posterior (median and 95\% CrI) survival time.
    The true survival time is shown in black.
    $\psensi$ is indicated on the panel labels.
    (A-C) $\psenss = 0.8$.
    (D-F) $\psenss = v$.
  }
  \label{imperf-test:fig:test-sensitivity}
\end{figure}

Next, we considered the consequence of $\psens$ being misspecified and using the strongly informative prior, the better performing prior in the correctly specified case.
If the test sensitivity is misspecified then the estimate of the survival distribution is biased.
If $\psensi < \psenss$, then the posterior estimate initially follows the true value but then separates (see \cref{imperf-test:fig:test-sensitivity}(A)).
The number of episodes inferred to have truly ended by the first negative is too low, and hence the survival function is overestimated.
This effect dominates over the opposing bias of overestimating the number of undetected episodes.
The opposite occurs if $\psensi > \psenss$, although the posterior moves away from the truth earlier (see \cref{imperf-test:fig:test-sensitivity}(C)).

The results when $\psenss = v$ are similar to $\psenss = 0.8$ (see \cref{imperf-test:fig:test-sensitivity}).
This suggests that the simplified model, with constant test sensitivity, is sufficient for recovering the true survival time.
Therefore, we conclude that including a varying test sensitivity is not required for adequate inference, and apply it to the real CIS data in the next section.
\begin{figure}
\end{figure}

\section{Application to the CIS data} \label{sec:CIS}

\begin{figure}
  \thisfloatpagestyle{empty}
  \makebox[\textwidth][c]{\includegraphics[width=0.9\paperwidth]{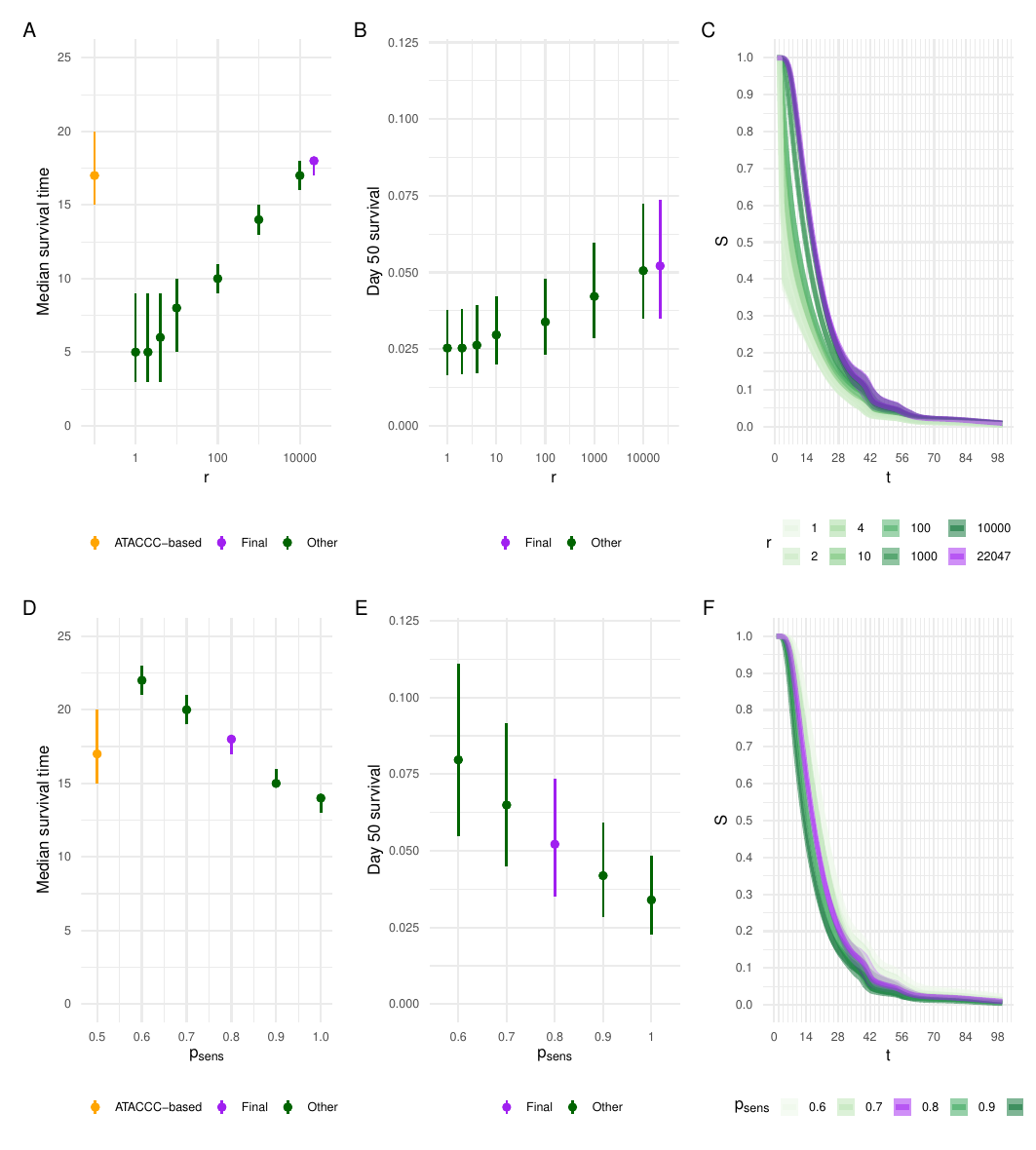}}
  \caption{%
    Assessing prior sensitivity.
    (A-C) Changing $r$ when $\psens = 0.8$.
    (D-F) Changing $\psens$ when $r = r\inform$.
    A and D: median survival time, compared to ATACCC-based estimate (shown in orange).
    B and E: $S_\theta(50)$.
    C and F: $S_\theta(t)$ for $t \in [1, 100]$.
    The final estimate is shown in purple throughout, green estimates are sensitivity analyses.
  }
  \label{imperf-test:fig:cis-sensitivity}
\end{figure}

In this section we apply the approach described in this chapter to the CIS infection episode dataset.
Unlike in the simulation studies, an uninformative prior on $\ntot$ led to implausible estimates of the duration distribution (small values of the prior's overdispersion parameter $r$ in \cref{imperf-test:fig:cis-sensitivity}(A) give estimates with a median survival time of 5 days).
The uninformative prior led to high posterior estimates of $\ntot$, and hence an implausibly large number of episodes with durations of less than five days.
Therefore, we used an informative prior for $\ntot$, $\ntot \sim \NBc(\mu\inform, r\inform)$ from pre-existing estimates of the total number of infections to give $\mu\inform$ and $r\inform$.
\citet{birrellRTM2} estimated the total number of infections in England over the time period we consider, with posterior mean \numprint{4136368} and standard deviation \numprint{27932}.
Approximating this distribution as a negative binomial and scaling the mean to the size of the CIS sample gives the prior $\mu\inform = 25132$ and $r\inform = 22047$.

\begin{figure}
  \centering \includegraphics{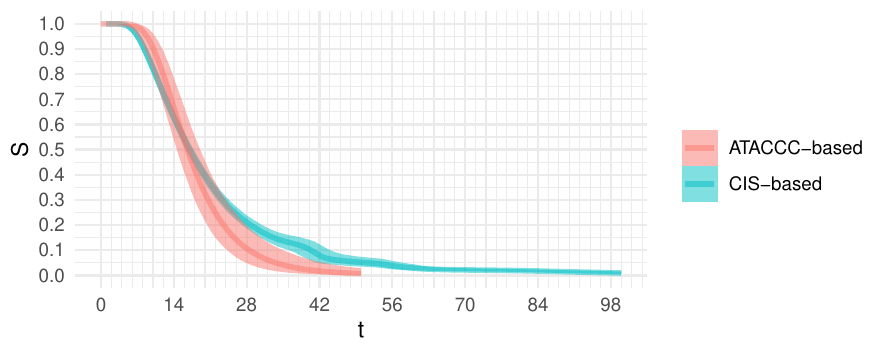}
  \caption{Duration estimates using CIS and ATACCC data}
  \label{imperf-test:fig:cis-estimates}
\end{figure}

With this prior, the model produces plausible estimates of the duration distribution (see \cref{imperf-test:fig:cis-estimates}).
At values above 50\%, the survival function (blue) decreases more slowly than the ATACCC-based estimate used for the prior in \cref{sec:ataccc-prior} (red), indicating a greater number of infections lasting much longer than the median.

The increase in long episodes (\eg longer than 50 days) is not very sensitive to the choice of prior for $\ntot$, the assumed value for $\psens$ (see \cref{imperf-test:fig:cis-sensitivity}), and the choice of prior for the hazards, $\lambda_t$.
However, the survival proportion over the first 4 weeks is sensitive to these choices.
The estimate using a test sensitivity of 0.8 and $\NBc(\mu\inform, r\inform)$ gives a median survival time most similar to the ATACCC-based estimate.

\begin{SCfigure}
  \centering \includegraphics{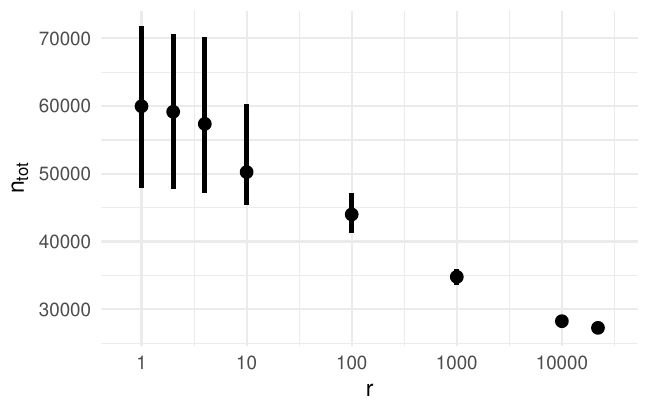}
  \caption[Sensitivity of $\ntot$'s posterior to its prior.]{How the posterior estimate of $\ntot$ changes with the value of $r$ in the prior on $\ntot$.}
  \label{imperf-test:fig:ntot}
\end{SCfigure}

Our estimates are sensitive to the choice of $r$, the strength of the prior on $\ntot$.
A low value for $r$, giving a very weak, almost uninformative, prior on $\ntot$ causes its posterior estimate to be much higher than the estimate from \citet{birrellRTM2}.
When increasing the prior's strength, the posterior estimate moves towards the prior smoothly, as expected (see \cref{imperf-test:fig:ntot}).
As discussed previously, the prior information is reliable for the first 2--3 weeks, notably including the median time.
The median using $r\inform$ matched the prior's median estimate and is a principled choice because it is based directly on the previous posterior estimate~\citet{birrellRTM2}.
Therefore, we recommend this estimate, which has a mean survival time of 21.2 days (95\% CrI: 20.5--21.9).

\section{Discussion} \label{sec:discussion}

This work is motivated by the challenge of exploiting data from the CIS, a unique long-running general population prevalence study conducted during the COVID-19 pandemic, to estimate the duration of SARS-CoV-2 infection episodes.
This is a key component in the estimation of incidence of infection and has an essential role in informing pandemic mitigation strategies such as isolation.

To estimate duration, we extended the survival analysis framework in \citet{heiseyModelling} to deal with the CIS design. The result is new methodology to analyze doubly censored data with imperfect test sensitivity and undetected events, when these undetected events occur with an arbitrary pattern in known individuals.
We estimate a nonparametric discrete-time survival distribution in a fully Bayesian framework and incorporate data from a complementary study, ATACCC, through an appropriately-discounted prior.

These CIS data are unique, but the study may serve as a template for studies in future pandemics~\citep{hallettModule}.
Our methodological framework is generic; therefore, it could be used to analyze these studies.
Furthermore, the simulation framework we developed can assist with designing more efficient studies, for example embedding an intensive, ATACCC-like study with a CIS-like study.

We estimate a mean duration of 21.2 days (95\% CrI: 20.5--21.9) with 5.3\% (95\% CrI: 3.5--7.4\%) of episodes lasting 50 days or longer.
The proportion of long episodes is higher than previous estimates (see \cref{tab:compare-survival}), leading to a longer mean.
However, previous studies did not accurately quantify the proportion of long episodes because of a lack of long-term follow-up and/or small sample size.
Additionally, CIS has a broader population base, including older individuals and those with more co-morbidities who may have longer durations.
Therefore, our estimates are more robust.
Crucially, the longer mean duration implies lower incidence when estimated from known prevalence.

\begin{table}[]
\centering
\begin{tabular}{llll}
Days from first positive & \citet{killingleySafety}  & ATACCC-based     & Ours                \\
12                       & 100\% (81--100\%) & 80\% (67--90\%)  & 72\% (69--75\%)     \\
26                       & 33\% (13--60\%)   & 14\% (7.1--25\%) & 24\% (22-27\%)      \\
88                       & 0\% (0--19\%)     & N/A              & 1.5\% (0.97--2.1\%)
\end{tabular}
\caption{
    Comparison of survival time estimates from
    \Citet{killingleySafety} (assuming a two day time from inoculation to positive with 95\% binomial confidence intervals using the Clopper--Pearson method~\citep{clopperUse}).
    ATACCC-based is from \cref{sec:ataccc-prior}; the analysis does not extend to 88 days.
    Ours is our final posterior (see \cref{sec:CIS}).
}
\label{tab:compare-survival}
\end{table}

Our methodology depends on several assumptions.
An important one is that infection episodes are independent, a simplification because infection episodes start times are correlated between different individuals.
During an epidemic, there are periods of time when all individuals are at higher risk of infection, due to the disease having a high prevalence in the population.
We mitigated this issue by choosing a period of time when prevalence was fairly constant.
In addition, we conducted a sensitivity analysis to the simulation study allowing underlying infection rates to either grow or shrink exponentially; these did not make a substantial difference to the estimates (not shown).
However, this non-independence is further complicated by the household structure of the CIS.
Individuals in the same household are likely to infect each other, and hence have clustered times at which their infection episodes begin.
Additionally, they have similar or the same testing schedules (due to the CIS study design).
The likely affect of this is that the uncertainty in our estimates is underestimated, although it should not introduce bias.
Further work could quantify the impact of this issue.

Infection episodes within the same individual also affect each other.
An infected individual who recovers is less likely to be infected in the future, due to having some immunity to the disease.
Additionally, the multinomial likelihood proposed in \cref{sec:inference} allows the possibility of concurrent infections; however, these would appear in the dataset as the same infection episode.
In contrast, the simulated data allowed each individual to experience at most one infection, which is likely as we consider a short period of time~\citep{milneImmunity}.
Despite this, in the simulation study without false negatives, the method we propose recovered the true survival function.
Therefore, it is unlikely that assuming independence of infection episodes in the same individual matters, possibly due to the large number of individuals without detected episodes.
Further work could explore alternative assumptions, \eg a ``full immunity'' assumption limiting each individual to one episode in the period.

Further assumptions were also required when extending the framework to allow false negatives.
Most importantly, that the negative immediately before a detected episode is a true negative, and that there is a negligible probability of missing an episode due to two false negatives.
Our simulation study shows that these assumptions are reasonable, and do not substantially impact performance when $\psens$, the test sensitivity, is high.
However, when $\psens$ is low, these can lead to biased results.
A promising direction for future work would look at the relaxation of these assumptions, for example, by inferring $\psens$ as a function of time since the episode began, similar to the generative model in \cref{sec:false-negatives}.

Ideally, $\psens$ would be estimated from the data.
However, this would require incorporating time-varying test sensitivity into the likelihood.
If the current model, with a constant $\psens$, is used then the estimate of $\psens$ would be heavily informed by intermittent negatives which will generally be close to the beginning of the episode with a higher test sensitivity than average.
Estimating the test sensitivity excluding intermittent negatives is not possible because the likelihood is monotonically decreasing in $\psens$; therefore, the likelihood always favors $\psens = 0$ (\ie no true positives).

An important avenue for further work is removing the need for an informative prior for $\ntot$, the total number of infections in the cohort in the period (including undetected infections).
The first step for such work would be to identify the reason an informative prior is required, for example through simulation studies with different patterns of false negatives.
We provide an R package implementing a flexible simulation framework for enabling this work.

A final challenge this study faced is the use of the SRS, which had limited computational power as well as lengthy approval processes for software or data to be moved in or out of the environment.
To enable inference in this low-resource environment, our method avoids increasing the dimensionality of the problem as far as possible.
Future pandemic plans should consider how to ensure privacy for study participants while allowing rapid data analysis and inter-operation.

\import{.}{supplemental}

\bibliographystyle{agsm}

\bibliography{references}

\end{document}

%% file: supplemental.tex
\def\spacingset#1{\renewcommand{\baselinestretch}%
{#1}\small\normalsize} \spacingset{1}

\appendix

\section{Derivations of quantities in \cref{sec:inference}} \label{sec:derivations}

\subsection{Expressions in \cref{sec:inference}}

First, the derivation of $p(\vec{\theta} \mid \na)$:
\begin{align}
p(\vec{\theta} \mid \na)
&\propto p(\vec{\theta}) p(\na \mid \vec{\theta}) \\
&= p(\vec\theta) \sum_{\ntot= \ndet}^{\infty} p(\ntot, \na \mid \vec{\theta}) \\
&= p(\vec{\theta}) \sum_{\ntot=\ndet}^\infty p(\ntot \mid \vec{\theta}) p(\na \mid \ntot, \vec{\theta}) \\
&= p(\vec{\theta}) \sum_{\ntot=\ndet}^\infty p(\ntot \mid \vec{\theta}) \frac{\ntot!}{(\ntot - \ndet)!} \pnodet^{\ntot - \ndet} \prod_{k \in \set{D}} p_k &\text{by \cref{perf-test:eq:multinomial}} \\
&= p(\vec{\theta}) \left( \prod_{k \in \set{D}} p_k \right) \left( \sum_{\ntot=\ndet}^\infty p(\ntot \mid \vec{\theta}) \frac{\ntot!}{(\ntot - \ndet)!} \pnodet^{\ntot - \ndet} \right).
\intertext{For convenience, define the summation term as:}
\eta &= 
\sum_{\ntot=\ndet}^\infty p(\ntot \mid \vec{\theta}) \frac{\ntot!}{(\ntot - \ndet)!} \pnodet^{\ntot - \ndet}. \label{perf-test:eq:eta}
\end{align}

Next, we derive an analytical solution to $\eta$ (defined in \cref{perf-test:eq:eta}) assuming the prior $\ntot \dist \NBc(\mu, r)$, and that it is independent of $\vec{\theta}$, the parameters of the survival distribution.
Therefore, $p(\ntot \mid \vec{\theta}) = p(\ntot)$.
This assumption makes $\eta$ analytically tractable, allowing computationally feasible inference.

Putting a negative binomial prior on $\ntot$ is equivalent to the following gamma-Poisson composite; its use simplifies the derivation.
\begin{align}
\ntot \mid \lambda &\dist \Poi(\lambda) \\
\lambda &\dist \GamDist(a, b)
\end{align}
where $b = r / \mu$ and $a = r$.
Hence:
\begin{align}
\eta
&= \int \sum_{\ntot=\ndet}^\infty \frac{\ntot!}{(\ntot-\ndet)!} \pnodet^{\ntot-\ndet} p(\ntot \mid \lambda) p(\lambda) d\lambda &\text{$\lambda$ explicit}\\
&= \int \sum_{\ntot=\ndet}^\infty \frac{\ntot!}{(\ntot-\ndet)!} \pnodet^{\ntot-\ndet} \frac{\lambda^{\ntot} e^{-\lambda}}{\ntot!} p(\lambda) d\lambda &\ntot \dist \Poi\\
&= \int \lambda^{\ndet} e^{-\lambda} p(\lambda) \sum_{\nnodet=0}^\infty \frac{(\pnodet \lambda)^{\nnodet}}{\nnodet!} d\lambda &\nnodet = \ntot-\ndet\\
&= \int \lambda^{\ndet} e^{-\lambda} p(\lambda) e^{\lambda \pnodet} d\lambda &\text{Maclaurin series of $e$} \\
&= \int \lambda^{\ndet} e^{-\lambda(1 - \pnodet)} \frac{b^a}{\Gamma(a)} \lambda^{a-1} e^{-b\lambda} d\lambda &\lambda \dist \GamDist\\
&= \int \frac{b^a}{\Gamma(a)} \lambda^{a+\ndet-1} e^{-(b+1-\pnodet)\lambda} d\lambda \\
&= \frac{b^a}{\Gamma(a)} \frac{\Gamma(a+\ndet)}{(b+1-\pnodet)^{a+\ndet}} &\text{Gamma pdf}\\
&\propto (b+1-\pnodet)^{-(a+\ndet)} &\text{only $p_u$ depends on $\theta$}\\
&= (r/\mu + 1 - \pnodet)^{-(r+\ndet)} &\text{sub in $\mu$ and $r$}\\
&\propto(r + \mu (1- \pnodet))^{-(r+\ndet)}.
\end{align}

\subsection{Expressions in \cref{sec:prob-undetected}}

If $i_j = i_k$ then the event $O_j = \vec{\nu}_k$ occurs if and only if the episode starts in the interval $[l^{(b)}_k, r^{(b)}_k]$ and ends in the interval $[l^{(e)}_k, r^{(e)}_k]$.
 on $\vec{\theta}$ and $i_j = i_k$, this gives:
\begin{align}
p_{ik}
=& \prob \left( l_k^{(b)} \leq B_{j} \leq r_k^{(b)}, l_k^{(e)} \leq E_{j} \leq r_k^{(e)} \right) \\
=& \prob \left( l_k^{(e)} \leq E_{j} \leq r_k^{(e)} \mid l_k^{(b)} \leq B_{j} \leq r_k^{(b)} \right) \times\prob \left( l_k^{(b)} \leq B_{j} \leq r_k^{(b)} \right) \\
=& \sum_{b = l_k^{(b)}}^{r_k^{(b)}} \prob \left( l_k^{(e)} \leq E_{j} \leq r_k^{(e)} \mid B_{j} = b \right) \prob \left(B_{j} = b \right) \\
=& \sum_{b = l_k^{(b)}}^{r_k^{(b)}} \prob \left( l_k^{(e)} - b + 1 \leq D_{j} \leq r_k^{(e)} - b + 1 \right) \prob \left(B_{j} = b \right) &\text{by def of $D_{j}$} \\
=& \sum_{b = l_k^{(b)}}^{r_k^{(b)}} \left( S_{\vec{\theta}}(l_k^{(e)} - b + 1) - S_{\vec{\theta}}(r_k^{(e)} - b + 2) \right) \prob \left(B_{j} = b \right) &\text{by def of $S_{\vec{\theta}}$} \\
\propto& \sum_{b = l_k^{(b)}}^{r_k^{(b)}} \left( S_{\vec{\theta}}(l_k^{(e)} - b + 1) - S_{\vec{\theta}}(r_k^{(e)} - b + 2) \right)
\end{align}
under the assumption of uniform probability of infection time.
This is the standard form of the likelihood for doubly interval censored data without truncation~\citep[e.g.][]{sunEmpirical}.

\subsection{Expression for $p_{u}$}

\begin{align}
  1 - p_u
  &= 1 - \sum_{i=1}^{\Ncis} \prob(O_j = \varnothing, i_j = i \mid \vec{\theta}) \\
  &= 1 - \sum_{i=1}^{\Ncis} \prob(O_j = \varnothing \mid i_j = i, \vec{\theta}) P(i_j = i \mid \vec{\theta}) \\
  &= 1 - \frac{1}{\Ncis}\sum_{i=1}^{\Ncis} \prob(O_j = \varnothing \mid i_j = i, \vec{\theta}) \\
  &= \frac{1}{\Ncis} \sum_{i=1}^{\Ncis} (1 - \prob(O_j = \varnothing \mid i_j = i, \vec{\theta}))
\end{align}

\section{Derivation of quantities in \cref{sec:false-negatives}}

\subsection{Expressions in \cref{imperf-test:sec:modifying-p_ia}} \label{sec:p-ia-dash}

We proceed by first considering whether $E_j > r_k^{(e)}$ is the case and conditioning on $B_j = b$.
Then, we combine the cases and remove the conditioning.

First, the case when $E_j \leq r_k^{(e)}$.
In this case, the test at $r_k^{(e)}+1$ is a true negative and the end of the episode is interval censored as in the previous chapter.
The true negative occurs with probability 1, by the assumption of no false positives.
\begin{align}
&\prob(O'_j = \vec{\nu}_k', E_j \leq r_k^{(e)} \mid B_j = b, i_j = i_k, \psens, \vec{\theta}) \\
&= \prob(O'_j = \vec{\nu}_k', l_k^{(e)} \leq E_j \leq r_k^{(e)} \mid B_j = b, i_j = i_k, \psens, \vec{\theta}) &\text{the test at $l_k^{(e)}$ is positive} \\
&= \prob(O'_j = \vec{\nu}_k' \mid l_k^{(e)} \leq E_j \leq r_k^{(e)}, B_j = b, i_j = i_k, \psens, \vec{\theta}) \\
&\ \ \  \times \prob(l_k^{(e)} \leq E_j \leq r_k^{(e)} \mid B_j = b, i_j = i_k, \psens, \vec{\theta}) \\
&= p_\text{sens}^{t_+} (1 - p_\text{sens})^{f_-} \left( S_{\vec{\theta}}(l_k^{(e)} - b + 1) - S_{\vec{\theta}}(r_k^{(e)} - b + 2) \right)
\label{imperf-test:eq:ll-ei-lt-ri}
\end{align}

Second, the case when $E_j > r_k^{(e)}$.
In this case, the test at $r_k^{(e)}+1$ is a false negative, occurring with probability $(1 - p_\text{sens})$.
To avoid having to consider tests after $r_k^{(e)}$, which could greatly complicate the likelihood, we model this case as the episode being right censored at $r_k^{(e)}$.
Taking the same approach as before:
\begin{align}
&\prob(O'_j = \vec{\nu}_k', E_j > r_k^{(e)} \mid B_j = b, i_j = i_k, \psens, \vec{\theta}) \\
&= \prob(O'_j = \vec{\nu}_k' \mid E_j > r_k^{(e)}, B_j = b, i_j = i_k, \psens, \vec{\theta}) \\
  &\ \ \  \times \prob(E_j > r_k^{(e)} \mid B_j = b, i_j = i_k, \psens, \vec{\theta}) \\
&= p_\text{sens}^{t_+} (1 - p_\text{sens})^{f_-} (1 - p_\text{sens}) S_{\vec{\theta}}(r_k^{(e)} - b + 2)
\label{imperf-test:eq:ll-ei-gt-ri}
\end{align}

These expressions can now be used to derive $p'_{ik}$.
First, augment the data with $b$, and split into the cases just discussed, omitting the conditioning on $\psens$, $\vec{\theta}$, and $i_j = i_k$:
\begin{align}
p_{ik}'
=& \prob(O'_j = \vec{\nu}_k') \\
=& \sum_{b = l_k^{(b)}}^{r_k^{(b)}} \left( \prob(O'_j = \vec{\nu}_k', E_j \leq r_k^{(e)} \mid B_j = b) + \prob(O'_j = \vec{\nu}_k, E_j > r_k^{(e)} \mid B_j = b) \right) \prob(B_j = b). \\
\intertext{Now, substitute in \cref{imperf-test:eq:ll-ei-lt-ri,imperf-test:eq:ll-ei-gt-ri} and take out the common factor:}
=\ &  p_\text{sens}^{t_+} (1 - p_\text{sens})^{f_-} \\
 & \times \sum_{b = l_k^{(b)}}^{r_k^{(b)}} \left( S_{\vec{\theta}}(l_k^{(e)} - b + 1) - S_{\vec{\theta}}(r_k^{(e)} - b + 2) + (1 - p_\text{sens}) S_{\vec{\theta}}(r_k^{(e)} - b + 2) \right) \\ 
  & \times \prob(B_j = b \mid p_\text{sens}, \vec{\theta}) \\
=\ &  p_\text{sens}^{t_+} (1 - p_\text{sens})^{f_-} \\
  & \times \sum_{b = l_k^{(b)}}^{r_k^{(b)}} \left( S_{\vec{\theta}}(l_k^{(e)} - b + 1) - p_\text{sens} S_{\vec{\theta}}(r_k^{(e)} - b + 2) \right) \\
  & \times \prob(B_j = b \mid p_\text{sens}, \vec{\theta}).
\label{imperf-test:eq:pia-prime}
\end{align}
Note that if $p_\text{sens} = 1$ then $p_{ik}' = p_{ik}$ (see \cref{perf-test:eq:pia}).

We use a fixed $\psens$ (\ie a point prior) and $\prob(B_j = b \mid \psens, \vec{\theta}) \propto 1$ giving:
\begin{align}
p_{ik}'
&\propto \sum_{b = l_k^{(b)}}^{r_k^{(b)}} S_{\vec{\theta}}(l_k^{(e)} - b + 1) - p_\text{sens} S_{\vec{\theta}}(r_k^{(e)} - b + 2).
\end{align}
Estimating $\psens$ is not possible in the current framework (see \cref{sec:discussion}).

\subsection{Expressions in \cref{imperf-test:sec:modifying-p_iu}} \label{sec:p-iu-dash}

The probability of one of the conditions that cause an episode to be missed (specified in \cref{imperf-test:sec:modifying-p_iu}) occurring, conditional on $B_j = b$ where $\min(\sched_{i_j}) < b \leq T_{i_j}$ is:
\begin{align}
&\prob \left(
    \tau_{\sched_{i_j}}(b) + 1 \leq D_j \leq \tau^2_{\sched_{i_j}}(b)
    \mid B_j = b, \vec{\theta} \right) (1 - \psens) \\
&= \left( S_{\vec{\theta}}(\tau_{\sched_{i_j}}(b) + 1) - S_{\vec{\theta}}(\tau^2_{\sched_{i_j}}(b) + 1) \right) (1 - \psens).
\end{align}
Summing over $b$, in the same way as \cref{perf-test:eq:piu}, gives:
\begin{align}
\zeta = (1 - p_\text{sens})\frac{1}{T} \sum_{b=\min(\sched_{i_j}) + 1}^{T_{i_j}} \left( S_{\vec{\theta}}(\tau_{\sched_{i_j}}(b) + 1) - S_{\vec{\theta}}(\tau^2_{\sched_{i_j}}(b) + 1) \right).
\label{imperf-test:eq:zeta}
\end{align}

$p_{iu}'$ is the probability of episode $i$ being undetected, considering both the previous and new mechanisms.
The previous and new mechanisms are mutually exclusive.
Hence, $p_{iu}'$ is the sum of these, $p_{iu}' = p_{iu} + \zeta$.
As previously, $1 - p_{iu}'$ is the required quantity.
\begin{align}
1 - p_{iu}'
=& 1 - p_{iu} - \zeta \\
=& \frac{1}{T} \sum_{b=\min(\sched_{i_j}) + 1}^{T_{i_j}} \left( p_\text{sens} S_{\vec{\theta}}(\tau_{\sched_{i_j}}(b) + 1) + (1 - p_\text{sens}) S_{\vec{\theta}}(\tau^2_{\sched_{i_j}}(b) + 1)\right).
\end{align}

\section{ATACCC-based prior} \label{sec:ataccc-prior}

Reliable estimates of $\lambda_t$ for $t$ up to around 20 are available from a prior analysis of data from The Assessment of Transmission and Contagiousness of COVID-19 in Contacts (ATACCC) study~\citep{hakkiOnset}, which tested individuals who had been exposed to infection daily up to a maximum of 20 days.
The infrequent testing in CIS means that there is a lack of information about short infection episodes, and hence we use these estimates as informative priors.

When constructing the prior, two aspects need consideration.
Firstly, the model structure from the ATACCC-based analysis leads to its posterior distribution having a positive correlation between $\lambda_t$ and $\lambda_{t'}$, especially for small $|t-t'|$.
The prior used in this analysis should preserve this correlation.
Secondly, the uncertainty in the prior estimates for $\lambda_t, t\geq20$ are underestimated because they are based on extrapolation of the ATACCC data using strong model assumptions.

We first approximate the previous posterior estimate of the hazard as $\logit{\vec{h}} \dist \MNorm(\vec{\mu}_A, \matr{\Sigma}_A)$ where $\vec{h}$ is the hazard, and $\vec{\mu}_A$ and $\matr{\Sigma}_A$ are the mean and covariance matrix estimated using samples of $\logit{\vec{h}}$ from the previous study's posterior.
Using a multivariate normal, as opposed to multiple univariate distribution for each $h_t$, preserves the correlation between the hazards.
The approximation is very good (not shown).

Having approximated the estimate as a multivariate normal, we add additional uncertainty using a discrete Beta process.
The discrete Beta process prior~\citep{ibrahimBayesian,sunStatisticala} generalises the form of prior used in the weakly informative case by allowing the central estimate of the hazard to vary over time.
It is:
\begin{align}
  \logit \vec{h} &\dist \MNorm(\vec{\mu}_A, \matr{\Sigma}_A) \\
  \lambda_t &\dist \text{Beta}(\alpha_t, \beta_t) &t = 1, 2, \dots \\
  \alpha_t &= k_t h_t + \alpha_0 \\
  \beta_t &= k_t (1 - h_t) + \beta_0
\end{align}
where $k_t$, $\alpha_0$, and $\beta_0$ are hyperparameters.
An intuition for what this distribution represents derives from a conjugate model for $\lambda_t$ with a beta prior and a binomial likelihood.
If $\lambda_t$ is given the prior distribution $\text{Beta}(\alpha_0, \beta_0)$, and we then have $k_t$ observations with $k_t h_t$ successes, then the posterior distribution for $\lambda_t$ is $\text{Beta}(\alpha_t, \beta_t)$ (as defined above).

$k_t$ reflects the subjective belief that $h_t$ is a good estimate of $\lambda_t$ for small $t$ but increasingly unreliable.
\begin{align}
k_t = \begin{cases}
  \expit(-0.4 * (t - 20)) &\text{for $t \leq 39$} \\
  0 &t > 39
\end{cases}.
\end{align}

\section{Distribution of duration used in simulation} \label{sec:simulation-truth}

The simulation requires a distribution of the duration of detectability.
We modify the ATACCC-based duration estimate from \citet[chapter 4]{blakeThesis} with an inflated tail to be consistent with the CIS.
The tail inflation uses a simple survival analysis and the CIS data.

This analysis assumes the initiating event is known, and equal to the episode’s detection time, $j_j^{(b)}$.
It assumes the final event is interval censored between the time of the final positive test and the subsequent negative test, or right censored if a negative test has not yet been observed.
A flexible, spline-based form is used for the baseline survival function~\citep{roystonSTPM,roystonFlexible} with covariates introduced via proportional odds.
By not accounting for either the undetected infections or the interval censoring of the initiating event, this analysis has competing biases which makes them hard to interpret~\citep{cisMethodsONS}.

To form the duration distribution used in the simulation, we combine the two estimates.
The pdf over the first 30 days is proportional to the ATACCC estimate, with the rest proportional to this CIS-based estimate.
Denote by $f_A(t)$ the ATACCC-based distribution function and $f_C(t)$ that from the CIS-based estimates just derived.
Then define:
$$
f_S'(t) = \begin{cases}
	f_A(t) &t \leq 30 \\
	f_C(t) &t > 30
\end{cases}
$$
Then the distribution used in the simulation is the normalised version of this: $f_S(t) = f'_S(t)/\sum_i f_S'(i)$.
Episode $j$'s duration of detectability is then a draw from this distribution.